\documentclass[11pt,a4paper]{article}
\usepackage{amsmath,amssymb,bbold,epsfig}
\usepackage{amsfonts}
\usepackage{slashed}
\usepackage{dsfont}
\usepackage{cite}
\usepackage{graphicx,color}
\usepackage{bbm}
\usepackage{bm}
\usepackage[normalem]{ulem}
\usepackage[usenames,dvipsnames]{xcolor}
\definecolor{lightgray}{gray}{0.90}
\usepackage[colorlinks=true,linkcolor=blue,citecolor=green,urlcolor=blue,
bookmarks=true,bookmarksopen=true,pdfpagemode=None,pdfstartview=FitH]{hyperref}

\newcommand{\be}{\begin{equation}}
\newcommand{\ee}{\end{equation}}
\newcommand{\bea}{\begin{eqnarray}}
\newcommand{\eea}{\end{eqnarray}}

\begin{document}

\title{\vspace{-2cm}
\vglue 1.3cm
\Large \bf
Quantum mechanics aspects and subtleties of \\
neutrino oscillations\,\footnote{Talk given at the 
{\em International Conference on the History of the Neutrino, Paris, France, 
September 5-7, 2018.}}}

\author{
{Evgeny~Akhmedov \thanks{Email: \tt
akhmedov@mpi-hd.mpg.de}
\vspace*{3.5mm}
} \\
{\normalsize\em
Max-Planck-Institut f\"ur Kernphysik, Saupfercheckweg 1,}\\
{\normalsize\em
69117 Heidelberg, Germany
\vspace*{0.15cm}}
}
\date{}

\maketitle 
\thispagestyle{empty} 
\begin{abstract} 
Neutrino oscillations appear to be a simple quantum mechanical phenomenon.
However, a closer look at them reveals a number of subtle points and
apparent paradoxes. Some of the basic issues of the theory of neutrino
oscillations are still being debated. I discuss, from a historical
perspective, how the role of quantum mechanical aspects of neutrino
oscillations was realized in the course of the development of the
theory of this phenomenon.
\end{abstract} 

\newpage

\tableofcontents

\section{\label{sec:nuQM}Neutrino oscillations and quantum mechanics}

I will discuss how the various quantum mechanical (QM) aspects of 
neutrino oscillations were realized in the course of the development of 
the theory of this phenomenon. I will also use this opportunity to 
revisit some subtle issues of this theory and discuss their resolution.

Neutrino oscillations is a periodic change of neutrino flavour. 
Particles usually change their identity in collisions with other 
particles or when they decay, so it may look strange that neutrinos can 
change their flavour without any external influence. However, the 
phenomenon of oscillations is actually well known in quantum mechanics.  
A textbook example is a 2-level quantum system. If one produces the 
system in one of its stationary states, $|\Psi_1\rangle$ or 
$|\Psi_2\rangle$, its time evolution is very simple: $|\Psi(t)\rangle 
=e^{-i E_1 t}|\Psi_1\rangle$ or $|\Psi(t)\rangle=e^{-i E_2 t}| 
\Psi_2\rangle$, respectively.
The probability for the system to remain in such a state does not change with 
time. However, if the system is prepared in a state that is a linear 
superposition of its stationary states,
\be
|\Psi(0)\rangle=a|\Psi_1\rangle+ b|\Psi_2\rangle\,,\qquad\quad(|a|^2+|b|^2=1)\,,
\label{eq:superp}
\ee
its time evolution is more complex:
\be
|\Psi(t)\rangle = a e^{-i\,E_1\, t}\,|\Psi_1\rangle + b\,e^{-i\,E_2\,t}\,
|\Psi_2\rangle\,.
\label{eq:timeev}
\ee
The probability that the system will be found in its initial state 
$|\Psi(0)\rangle$ at time $t$ is 
\bea
P_{\rm surv}=|\langle \Psi(0)|\Psi(t)\rangle|^2=
\left||a|^2\,e^{-i\,E_1\,t}+|b|^2\,
e^{-i\,E_2\,t}\right|^2 \qquad \\
= 1 - 4 |a|^2 |b|^2
\sin^2 [(E_2-E_1)\,t/2]\,.
\label{eq:Psurv}
\eea
It oscillates with time with the frequency $(E_2-E_1)/2$ and the 
amplitude that takes its maximum value when $|a|=|b|$ (``maximum 
mixing'') and vanishes when either $a$ or $b$ is zero (no mixing). This 
analogy gives a rather accurate description of the physical essence of 
neutrino oscillations, as in the charged-current processes neutrinos are 
produced (and detected) as flavour eigenstates, which are non-trivial 
linear superpositions of the eigenstates of free propagation (mass 
eigenstates). The above description in terms of evolution in time of a 
superposition of stationary states was actually used in most of the 
early papers on neutrino oscillations~\cite{P1,P2,P3,P4,P5}. It, 
however, leaves out the question of how neutrino flavour changes with 
distance traveled by the neutrinos (see the discussion of the ``time to 
space conversion'' procedure below, though).

\subsection{\label{tricky}Tricky issues}

Alhough neutrino oscillations appear to be a simple QM phenomenon, a closer 
look at them reveals a number of subtle points and apparent paradoxes. 
A number of fundamental issues of the theory of neutrino oscillations have 
been actively debated ever since the idea of the oscillations was put forward 
by Pontecorvo~\cite{Po1,Po2}\footnote{
The papers debating the basics of the neutrino oscillation theory 
are too plentiful to be cited here. An incomplete (but representative) list of 
references can be found on slide 5 of the presentation slides of this talk at 
\url{http://neutrinohistory2018.in2p3.fr/programme.html}. 
}.
These include
\begin{itemize}

\item
Do neutrino mass eigenstates composing a given flavour eigenstate 
have same energy or same momentum? 
 
\item
Can one use plane waves or stationary states for describing neutrino 
oscillations?

\item
Do the oscillations contradict energy-momentum conservation?

\item
Under what conditions can the oscillations be observed? 

\item
Is wave packet approach necessary for describing neutrino oscillations?

\item
When are the oscillations described by a universal (i.e.\ production-- \,and 
detection--independent) probability?

\item
Is the standard oscillation formula correct? What is its domain of 
applicability? 

\item
How to get correctly normalized oscillation probabilities? 

\item
Are the oscillation probabilities Lorentz invariant? 

\item
Why do we say that charged leptons are produced as mass eigenstates and
neutrinos as flavour states and not the other way around?

\item
Do neutrinos produced in $\pi\to l \nu_l$ \,decays oscillate when the charged 
lepton is not detected?

\item
Do charged leptons oscillate?

\end{itemize}

\noindent
I will now briefly discuss how quantum mechanics allows us to answer these 
questions. 

\section{\label{sec:master}Master formula for the probability of neutrino 
oscillations in vacuum}

In the standard approach to neutrino oscillations the state vector describing 
a flavour eigenstate neutrino $\nu_\alpha$ ($\alpha=e, \mu, \tau, ...)$ 
is considered to be a linear superposition of the state vectors of the mass 
eigenstate neutrinos $\nu_i$ ($i=1, 2, 3, ...$):  

\be
|\nu_\alpha\rangle ~=~
\sum_i U_{\alpha i}^*\,|\nu_i\rangle\,,
\label{eq:mix}
\ee
where $U$ is the leptonic mixing matrix. From this expression one can derive 
the master formula for the probability of $\nu_\alpha\to\nu_\beta$ 
oscillations in vacuum: 
\be
P_{\alpha\beta}(L) = 
\bigg|\sum_i U_{\beta i}^{}\; e^{-i \frac{\Delta
m_{ik}^2}{2p} L} \;U_{\alpha i}^*\bigg|^2. 
\label{eq:master1}
\ee
How is it usually obtained?

\subsection{\label{sec:sameEsamep}Simplified derivations: Same energy and same 
momentum approaches}

Derivations of the oscillation probability which can be found in many 
texts typically proceed as follows. 
The state vector describing a flavour eigenstate neutrino $\nu_\alpha$ 
produced at time $t=0$ and coordinate $\vec{x}=0$ is taken to be 
given by Eq.~(\ref{eq:mix}). Assuming that neutrinos are described by plane 
waves, the evolved neutrino state after time $t$ at the position $\vec{x}$ 
is then 
\be
|\nu(t,\vec{x})\rangle ~=~ \sum_i U_{\alpha i}^*\,e^{-i p_i x}
|\nu^{\rm mass}_i\rangle\,.
\label{eq:evolved1}
\vspace*{-2mm}
\ee
That is, each mass eigenstate $\nu_i$ picks up the phase factor 
$e^{-i \phi_i}$, 
where 
\be
\phi_i \,\equiv\, p_i\, x \,=\, E_i t-\vec{p}_i\,\vec{x}\,.
\label{eq:phase}
\ee
Projecting the evolved neutrino state on the flavour eigenstate $\nu_\beta$ 
and taking the squared modulus yields the oscillation probability 
\be
P(\nu_\alpha\to\nu_\beta;\,t,\vec{x}) \,=\, \big|\langle\nu_\beta
|\nu(t,\vec{x})\rangle\big|^2\,.
\label{eq:prob1}
\ee
To evaluate it, one needs to calculate the phase differences between 
different mass eigenstates (the oscillation phases) $\Delta\phi_{ik}$: 
\be
\Delta \phi\,=\, \Delta E \cdot t-\Delta \vec{p}\cdot \vec{ x}\,.
\label{eq:phase1}
\ee
Here the subscripts $ik$ are omitted from $\Delta\phi$, $\Delta E$ and 
$\Delta\vec{p}$ in order to simplify the notation.
Clearly, different neutrino mass eigenstates composing a given flavour state 
cannot simultaneously have the same energy and the same momentum, as otherwise 
they would have had the same mass. Therefore in many studies two simplified 
approaches were adopted: 

{\em (a) Same momentum approach.}  
Assume that all the mass eigenstates composing the produced neutrino flavour 
state have the same momentum, i.e.\ $\Delta\vec{p}=0$. Then 
Eq.~(\ref{eq:phase1}) gives $\Delta \phi=\Delta E \cdot t$, and the 
oscillation probability (\ref{eq:prob1}) depends only on the evolution time 
$t$. Since for ultra-relativistic neutrinos 
$E_i=\sqrt{\vec{p}\,^2+m_i^2}\simeq 
p+\frac{m_i^2}{2p}$, for the oscillation phase one finds 
\be
\Delta\phi
=\Delta E\cdot t
\simeq \frac{\Delta m^2}{2p}t\,.
\label{eq:phase2}
\ee 
Experimentally, the distance between the neutrino source and detector 
$L=|\vec{x}|$ rather than time of flight $t$ is normally known. 
It is then usually argued that, as neutrinos propagate with nearly the speed 
of light, 
\be
L\simeq t\,,
\label{eq:ttoL}
\ee
and so one can replace $t \to L$ in Eq.~(\ref{eq:phase2}) 
(in the literature on neutrino oscillations this procedure is sometimes called 
``time to space conversion''). With this replacement Eq.~(\ref{eq:phase2}) 
yields the usual 
oscillation phase, and using it in 
Eq.~(\ref{eq:prob1}) leads to the standard oscillation probability 
(\ref{eq:master1}).  
 
{\em (b) Same energy approach.}  
Assume now that all the mass eigenstates composing the produced neutrino 
flavour state have the same energy, i.e.\ $\Delta E=0$. Eq.~(\ref{eq:phase1}) 
then gives $\Delta \phi=-\Delta\vec{p} \cdot\vec{x}$. 
Assuming $\vec{x}||\vec{p}$ (which is well justified 
when the distance between the neutrino source and detector is large compared 
to their transverse sizes) one finds that  the oscillation probability 
(\ref{eq:prob1}) depends only on the distance $L$.  
For ultra-relativistic neutrinos one has 
$p_i=\sqrt{E^2-m_i^2}\simeq E-\frac{m_i^2}{2E}$, and the oscillation phase is 
\be
\Delta\phi=-\Delta p\cdot L \simeq \frac{\Delta m^2}{2E}L\,.
\label{eq:phase3}
\ee
This is the standard expression for the oscillation phase, which depends on 
the distance $L$ traveled by neutrinos; unlike in the ``same momentum'' 
approach discussed above, it was not necessary to invoke the ``time to space 
conversion'' procedure to arrive at it. The resulting oscillation probability 
is again that of Eq.~(\ref{eq:master1}). 

The above two approaches are very simple and transparent, and allow one to 
quickly get the desired result. The trouble with them is that they are both 
wrong. 

The point is that there is no reason whatsoever to expect the 
neutrino mass eigenstates composing a flavour state to have either the 
same energy or the same momentum. These assumptions actually contradict 
energy-momentum conservation. This was first demonstrated by 
R.~Winter~\cite{winter}, who considered neutrino emission in orbital 
electron capture by nuclei -- a process with 2-body final state and 
simple kinematics. Another process with 2-body final state -- charged 
pion decay -- was discussed in this context by Giunti and 
Kim~\cite{giunti1}. Let us follow their argument.

For a $\pi\to\mu\nu$ decay at rest, 4-momentum conservation gives for 
the energy and momentum of the produced 
neutrino mass eigenstate $\nu_i$ with mass $m_i$ 
\bea
E_i^2~=~\frac{m_\pi^2}{4}\left(1-\frac{m_\mu^2}{m_\pi^2}\right)^2
+\frac{m_i^2}{2}\left(1-\frac{m_\mu^2}{m_\pi^2}\right)
+\frac{m_i^4}{4m_\pi^2}\,,
\label{eq:Ei}
\\
p_i^2~=~\frac{m_\pi^2}{4}\left(1-\frac{m_\mu^2}{m_\pi^2}\right)^2
-\frac{m_i^2}{2}\left(1+\frac{m_\mu^2}{m_\pi^2}\right)
+\frac{m_i^4}{4m_\pi^2}\,.\,
\label{eq:pi}
\eea
Neglecting terms of order $m_i^4$, one finds  
\be
E_i~\simeq~E+\xi\frac{m_i^2}{2E}\,,\qquad
p_i~\simeq~E-(1-\xi)\frac{m_i^2}{2E}\,,
\label{eq:leading}
\ee
where 
\be
E\equiv \frac{m_\pi}{2}\left(1-\frac{m_\mu^2}{m_\pi^2}\right)\simeq 
30~{\rm MeV}\,,\qquad\quad
\xi~\equiv~\frac{1}{2}\left(1-\frac{m_\mu^2}{m_\pi^2}\right)~\approx~0.2. 
\label{eq:Exi}
\ee
As can be seen from Eq.~(\ref{eq:leading}), same energy and same momentum 
assumptions correspond to $\xi=0$ and $\xi=1$, respectively; in reality, 
however, $\xi$ is neither 0 nor 1 but about 0.2. Moreover, if we considered 
the $\pi\to e\nu$ decay rather than the $\pi\to\mu\nu$ one, we would have to 
replace $m_\mu$ by $m_e$ 
in Eqs.~(\ref{eq:Ei})-(\ref{eq:Exi}); for the parameter $\xi$ this would give  
$\xi\approx 0.5$, which is just in the middle between the values corresponding 
to same energy and same momentum assumptions. 

\subsection{\label{sec:moreproblems}Same $E$ and same $p$ approaches: More 
problems} 

The internal inconsistencies of the ``same energy'' and ``same momentum'' 
approaches to neutrino oscillations can actually be seen even without invoking 
energy-momentum conservation. In the same momentum approach is is assumed that 
neutrinos have well-defined momentum, i.e.\ they are described by plane waves. 
However, the probability to find a particle described by a plane wave has 
no coordinate dependence, i.e.\ it is the same at any point in space. This 
means that propagation of neutrinos in space cannot be accounted for in this 
case. For neutrino oscillation experiments it is crucial that neutrinos 
are produced and detected in distinct regions of space, the distance between 
which is the experimental baseline $L$. However, in the plane-wave approach 
one cannot even define the neutrino production and detection regions. 

Moreover, the oscillation phase in this case depends only on time 
(see Eq.~(\ref{eq:phase2})). Taken at face value, this result would lead to an 
absurd conclusion that in order to observe neutrino oscillations e.g.\ in 
reactor or accelerator neutrino experiments one would not need far detectors 
at all -- it would be sufficient to put the detector immediately next to the 
neutrino source and just wait long enough. 

In order to solve this problem, the ``time to space conversion'' procedure of 
Eq.~(\ref{eq:ttoL}) is usually invoked. Let us look at this procedure more 
carefully. One usually tries to justify it by the fact that neutrinos are 
ultra-relativistic, i.e.\ they propagate with nearly the speed of light.  
However, this is not the most important assumption behind Eq.~(\ref{eq:ttoL}). 
The same argument could be made even for non-relativistic neutrinos,  
provided that the different mass eigenstates composing a given flavour state 
move with nearly the same speed $v\ll 1$ (for which they would have to be 
nearly degenerate in mass). In that case one would merely have to replace 
Eq.~(\ref{eq:ttoL}) by $L\approx vt$, and the rest of the derivation of the 
oscillation probability would be essentially the same. What is much more 
important is that Eq.~(\ref{eq:ttoL}) (as well as its modified version 
$L\approx vt$) is only valid for {\em point-like particles moving along 
classical trajectories}. But the notion of a point-like particle is just the 
opposite of that of a plane wave! So, one tries to combine two incompatible 
approaches in this case. 

Similarly, same energy assumption based on the evolution of neutrino flavour 
only in space cannot account for the fact that neutrinos are produced and 
detected at certain \vspace*{1.3mm}times.  
 
So, the question is: How can two different and wrong assumptions (same $E$ and 
same $p$) result in the same (and correct) expression for the neutrino 
oscillation probability? To answer this question, it is necessary to consider  
a wave packet approach to neutrino oscillations. 

\section{\label{sec:WP}Wave packet approach: The basics}

In quantum theory localized particles are described by wave packets: instead 
of a plane wave one considers superpositions of plane waves with a 
momentum spread $\sigma_p$ around a central momentum $\vec{p}_0$. This is a 
consequence of the Heisenberg uncertainty relation: a state localized within a 
spatial region $\sigma_x$ is characterized by a momentum uncertainty 
$\sigma_p\gtrsim 1/\sigma_x$. Therefore it cannot be described by a plane 
wave,  for which $\sigma_p=0$. 
\begin{figure}[h]
\begin{minipage}{0.45\linewidth}
\centerline{\includegraphics[width=0.7\linewidth]{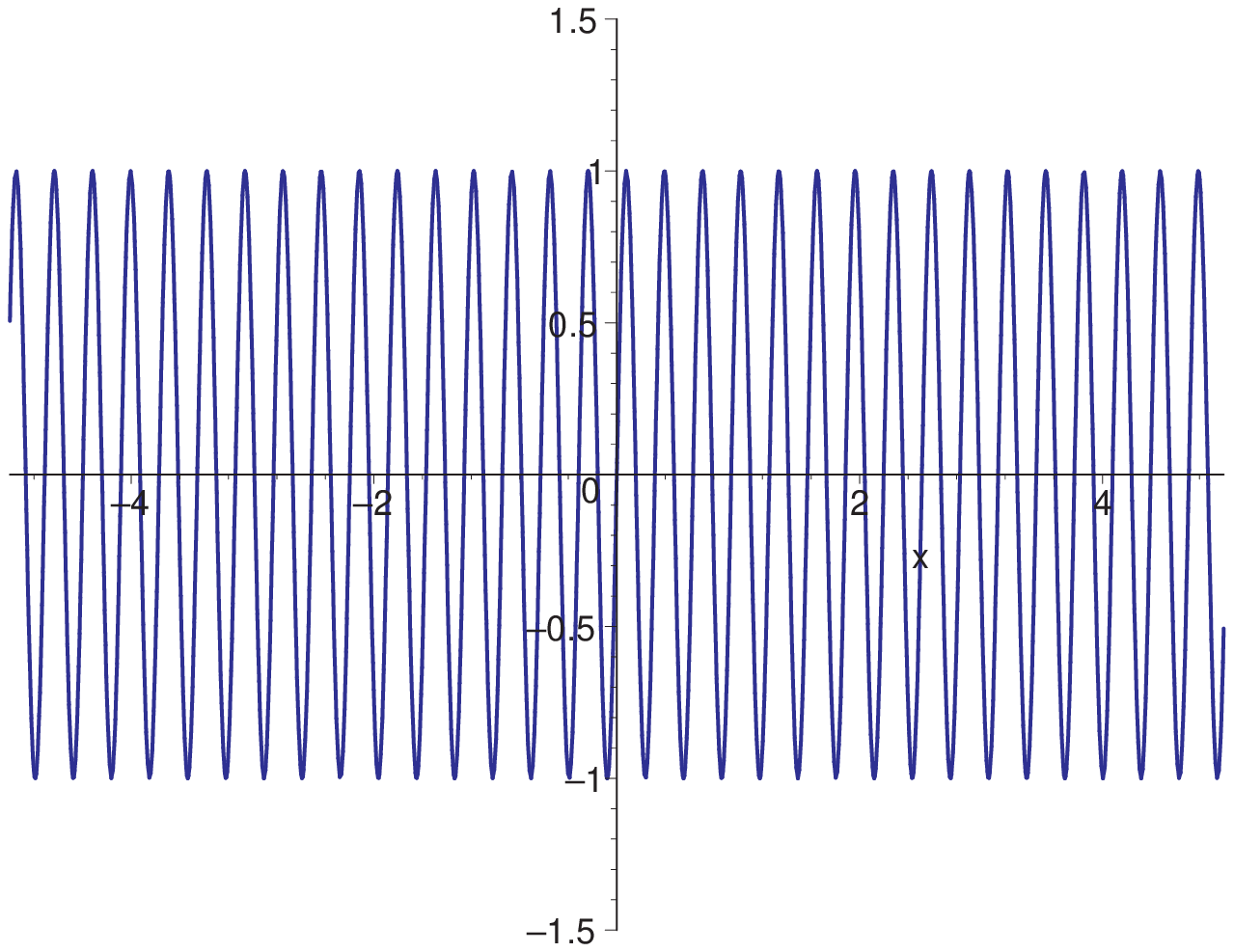}}
\end{minipage}
\hfill
\begin{minipage}{0.45\linewidth}
\centerline{\includegraphics[width=0.7\linewidth]{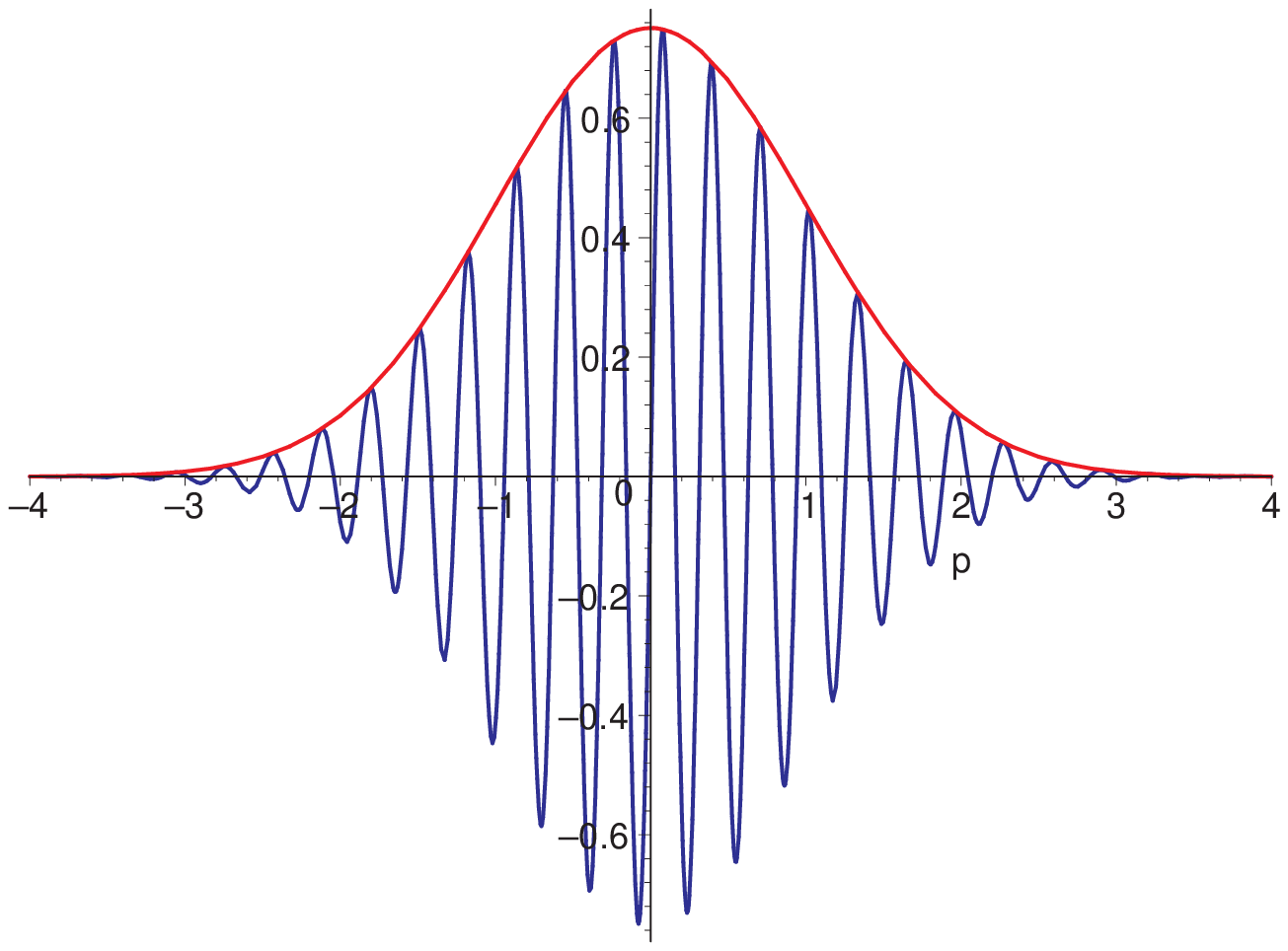}}
\end{minipage}
\hfill
\caption[]{Schematic representation of plane wave (left panel) and 
wave packet (right panel). 
}
\label{fig:wp}
\end{figure}
\noindent
In the wave packet approach the coordinate-space wave function of a free 
particle of mass $m_i$ is 
\be
\Psi_i(\vec{x},\, t)~=
\int\!\frac{d^3 p}{(2\pi)^{3}}\, 
f_{\vec{p}_0}(\vec{p})
\,e^{i \vec{p} \vec{x} - i E_i(p) t}\,,
\label{eq:wp1}
\ee
where $f_{\vec{p}_0}(\vec{p})$ is the momentum distribution amplitude with a 
peak at $\vec{p}=\vec{p}_0$ and a momentum width $\sigma_p$, and 
$E_i=\sqrt{\vec{p}\,^2+m_i^2}$. A frequently used example is the Gaussian 
wave packet: 
\be
f_{\vec{p}_0}(\vec{p})=\frac{1}{(2\pi\sigma_p^2)^{3/4}}
\exp\left\{-\frac{(\vec{p}-\vec{p}_0)^2}{4 \sigma_p^2}
\right\}.
\label{eq:gauss1}
\ee
If one neglects the spreading of the wave packet, the coordinate-space wave 
function of such a state takes the form 
\be
\Psi_i(\vec{x},\, t)=e^{i \vec{p}_0 \vec{x}-i E_i(p_0)t} \,
~\frac{1}{(2\pi\sigma_x^2)^{3/4}}\exp\left\{-\frac{(\vec{x}-\vec{v}_{gi} t)^2}
{4 \sigma_x^2}\right\}\,, 
\label{eq:wp}
\ee
where $\vec{v}_{gi}\equiv [\partial E_i(p)/\partial\vec{p}]_{\vec{p}=\vec{p}_0}
=\vec{p}_0/E_i(p_0)$ is the group velocity of the wave packet. Eq.~(\ref{eq:wp})
represents the plane wave corresponding to the central momentum $\vec{p}_0$ 
modulated by the Gaussian coordinate-space envelope factor of the width 
$\sigma_x=1/(2\sigma_p)$ (see Fig.~\ref{fig:wp}). The quantity $\sigma_x$ is 
therefore the spatial length of the wave packet. Note that though the time 
$t$ and the spatial coordinate $\vec{x}$ are, strictly speaking, independent, 
the probability of finding the particle $|\Psi_i(\vec{x},t)|^2$, 
which reaches its maximum at $\vec{x}=\vec{v}_{gi}t$,  
quickly decreases when $|\vec{x}-\vec{v}_{gi}t|$ starts exceeding 
$\sigma_x$. This holds true for the wave function of any localized state, 
not just for Gaussian wave packets. 

In the wave packet approach the evolved state for a neutrino which was 
produced as $\nu_\alpha$ is 
\be
|\nu(\vec{x},t)\rangle ~=~ \sum_i U_{\alpha i}^*|\nu_i(\vec{x},t)\rangle
~=~ \sum_i U_{\alpha i}^*
\,\Psi_i(\vec{x},t)
|\nu_i\rangle\,,
\label{eq:evolved2}
\vspace*{-1.5mm}
\ee
where the wave function $\Psi_i(\vec{x}, t)$ of the $i$th neutrino mass 
eigenstate is given in Eq.~(\ref{eq:wp1}). This is to be compared with the 
expression for the evolved neutrino state in the plane wave 
approach~(\ref{eq:evolved1}). The state of the detected neutrino $\nu_\beta$, 
on which the evolved state has to be projected in order to find the transition 
amplitude, should also be described by a localized wave packet. 
The wave packet approach to neutrino oscillations allows one to resolve many 
paradoxes and confusions in the oscillation theory.

\subsection{\label{sec:oscPhase}Oscillation phase and the wave packet approach}

How can one calculate the oscillation phase (\ref{eq:phase1}), now that we know 
that both same energy and same momentum approaches are actually incorrect? 
Let us take into account that we usually deal only with highly relativistic 
neutrinos.\footnote{The following arguments also apply to moderately 
relativistic as well as to non-relativistic neutrinos provided that they are 
nearly degenerate in mass.}
In this case the energy and momentum differences of the different neutrino mass 
eigenstates composing a given flavour state are small compared to their 
respective average values ($\Delta E \ll E$, $\Delta p \ll p$). One can 
therefore expand $\Delta E$ as 
\be
\Delta E=
\frac{\partial E}{\partial p} \Delta p + \frac{\partial E}
{\partial m^2} \Delta m^2 = v_g \Delta p  + \frac{1}{2E}\,\Delta m^2
\,,
\label{eq:expansion1}
\ee
where $v_g$ is the average group velocity of the neutrino mass eigenstates. 
Substituting this into Eq.~(\ref{eq:phase1}) yields 
\be
\Delta \phi
=-(L-v_g t) \Delta p + \frac{\Delta m^2}{2E}\,t\,.
\label{eq:phase4}
\ee
Let us examine this expression. If one adopts the (incorrect) same momentum 
approach, $\Delta p=0$, the first term on the right hand side (r.h.s.) 
vanishes, and the oscillation phase (\ref{eq:phase2}) is recovered. Note, 
however, that the first term on the r.h.s.\ of (\ref{eq:phase4}) vanishes also 
when $\Delta p\ne 0$ provided that $L=v_g t$, which corresponds to the center 
of the neutrino wave packet. Away from the center, $L-v_gt$ does not vanish 
but, as was discussed above, its value cannot significantly exceed the spatial 
length of the wave packet $\sigma_x$. Therefore, the first term on the r.h.s.\ 
is negligibly small and Eq.~(\ref{eq:phase2}) obtains without the unphysical 
``same momentum'' prescription provided that  
\be
\sigma_x\cdot |\Delta p|\ll 1\,.   
\label{eq:cond1}
\ee 
By the Heisenberg uncertainty relation $\sigma_x\sim 1/\sigma_p$, 
and therefore Eq.~(\ref{eq:cond1}) is equivalent to 
\be
|\Delta p|\ll \sigma_p\,.   
\label{eq:cond2}
\ee 
In addition, from $|L-v_g t|\lesssim \sigma_x$ it follows that one can replace 
$t\to L/v_g$ in the second term on the r.h.s of Eq.~(\ref{eq:phase4}) provided 
that $\sigma_x$ is negligibly small compared to the neutrino oscillation 
length. This yields the standard oscillation phase 
$\Delta\phi=[\Delta m^2/(2p)]L$. 

Quite similarly one can show that the correct oscillation phase leading 
to the standard oscillation probability (\ref{eq:master1}) can be obtained 
without the ``same energy'' assumption. Expressing $\Delta p$ from 
Eq.~(\ref{eq:expansion1}) and substituting it into Eq.~(\ref{eq:phase1}), one 
finds for the oscillation phase 
\be
\Delta \phi=-\frac{1}{v_g}(L~-~v_g\,t) \Delta E ~+~\frac{\Delta m^2}{2p}\,L\,, 
\label{eq:phase5}
\ee
which is equivalent to Eq.~(\ref{eq:phase4}). 
In the limit $\Delta E\to 0$ the first term on the r.h.s.\ of this equation 
vanishes and the results of the ``same energy'' approach are recovered. 
However, this term can also be neglected even when $\Delta E\ne 0$, provided 
that $(\sigma_x/v_g)|\Delta E|\ll 1$. As the spatial length of the wave packet 
satisfies $\sigma_x\sim 1/\sigma_p\simeq v_g/\sigma_E$ where $\sigma_E$ is 
the energy uncertainty of the neutrino state,\footnote{The relation 
$\sigma_E\simeq v_g\sigma_p$ follows from the mass-shell condition 
$E^2=\vec{p}\,^2+m^2$.} we find that the first term on the r.h.s.\ of 
Eq.~(\ref{eq:phase5}) can be neglected when 
\be
|\Delta E| \ll \sigma_E\,.
\label{eq:cond3}  
\ee

We can now answer the question raised at the end of 
Section~\ref{sec:sameEsamep}. Wrong ``same energy'' and ``same momentum'' 
assumptions lead to the correct oscillation probability (\ref{eq:master1}) 
because   
\begin{itemize}
\item
Neutrinos are relativistic 
with $|\Delta E| \ll E$, $|\Delta p| \ll p$, so that the expansion in 
Eq.~(\ref{eq:expansion1}) is justified. 

\item
In most (if not all) situations of practical interest, energy and momentum 
differences $\Delta E$ and $\Delta p$ 
are small compared to the intrinsic QM energy and momentum uncertainties of 
the neutrino state, i.e.\ conditions (\ref{eq:cond2}) and (\ref{eq:cond3}) 
are satisfied. 
\end{itemize}
As we shall see, Eqs.~(\ref{eq:cond2}) and (\ref{eq:cond3}) are essentially 
the coherence conditions for neutrino production and detection. 

\section{\label{sec:cohCond}When are neutrino oscillations observable? 
The role of QM uncertainties}

Quantum mechanics tells us that no particle can have precisely defined values 
of energy and momentum  -- these quantities always have some intrinsic 
uncertainties. This is due to the fact that particles are always localized 
in space and time. In particular, the processes in which the particles are 
produced and which actually determine their properties are always confined to 
finite space-time intervals. The QM uncertainty principles relate the momentum 
and energy uncertainties of a particle, $\sigma_p$ and $\sigma_E$, to the 
spatial extension and the duration of 
its production process, $\sigma_X$ and $\sigma_t$:\footnote{Strictly speaking, 
the QM uncertainty relations read  
$\sigma_p\ge (2\sigma_X)^{-1}$, $\sigma_E\gtrsim (\sigma_t)^{-1}$, that is, 
Eq.~(\ref{eq:uncert}) should actually contain inequalities rather than the 
$\sim$ symbol. However, except in very special cases of little practical 
interest, the relations in Eq.~(\ref{eq:uncert}) apply.}
\be
\sigma_p\sim \sigma_X^{-1}\,,\qquad\quad
\sigma_E\sim \sigma_t^{-1}\,.
\label{eq:uncert}
\ee
Usually, energy and momentum uncertainties of the particles are extremely 
small compared to their energies and momenta themselves; therefore, in most 
situations these uncertainties can be safely neglected. This is, however, not 
justified when neutrino oscillations are considered. The reason is that the 
neutrino energy and momentum uncertainties, as tiny as they are, are crucially 
important for the oscillation phenomenon -- without them the oscillations just 
would not occur. Indeed, as discussed above, if neutrinos were produced, for 
example, with no momentum uncertainty, this would mean that their source was 
completely delocalized in space, and therefore neutrino oscillations as a 
function of the distance $L$ between the neutrino source and detector would 
be unobservable. Similar arguments apply to the neutrino energy uncertainty. 

\subsection{\label{sec:cohproddet}Coherence conditions for neutrino 
production and detection}

The fact that too accurate measurement of neutrino energy and momentum would 
destroy neutrino oscillations was first demonstrated by Kayser~\cite{kayser}. 
Assume that by measuring the energies and momenta of all particles 
participating in a neutrino production process we reconstructed the energy $E$ 
and momentum $p$ of the produced neutrino with some accuracy. The errors 
in the determination of $E$ and $p$ cannot be smaller than the intrinsic QM 
uncertainties $\sigma_E$ and $\sigma_p$ related to the space-time localization 
of the production process. Assuming that these uncertainties are independent, 
from the mass-shell relation $E^2=\vec{p}\,^2+m^2$ one can then infer the 
squared mass of the emitted neutrino with the minimum uncertainty 
\be
\sigma_{m^2}=[(2E\sigma_E)^2+(2p\sigma_p)^2]^{1/2}\,.
\label{eq:sigmam}
\ee
If this minimum uncertainty is large compared to the difference between the 
squared masses of different neutrino mass eigenstates, i.e.\ $\sigma_{m^2}\gg 
|\Delta m^2|$, it is in principle impossible to find out which neutrino mass 
eigenstate was produced. This means that different neutrino mass eigenstates 
are produced coherently; what is actually emitted is a flavour eigenstate 
(\ref{eq:mix}), which is a coherent linear superposition of the neutrino mass 
eigenstates. 

Conversely, for $\sigma_{m^2}\lesssim |\Delta m^2|$ one can find out which 
neutrino mass eigenstate has actually been produced; this means that
different mass eigenstates cannot be produced coherently. As neutrino
oscillations are a result of interference of the amplitudes corresponding to
different neutrino mass eigenstates, the absence of their coherence means 
that no oscillations will take place. The flavour transition probabilities 
will then correspond to averaged neutrino oscillations. 

Coherence conditions for neutrino production and detection processes can also 
be formulated in configuration space. It was demonstrated by 
Kayser~\cite{kayser} that as soon as $\sigma_E$ and $\sigma_p$ become 
sufficiently small to allow determination of the neutrino mass at neutrino 
production, the uncertainty in the coordinate of the neutrino production point 
becomes larger than the oscillation length $l_{\rm osc}=4\pi p/\Delta m^2$, 
and so the oscillations get washed out due to the averaging over this 
coordinate. Similar arguments apply to neutrino detection. 
Thus, the production and detection processes cannot discriminate between 
different neutrino mass eigenstates (which is a necessary condition 
for the observability of neutrino oscillations) only when 
the uncertainties in the neutrino emission and absorption coordinates 
are sufficiently small, i.e.\ when the processes of neutrino production and 
detection are sufficiently well localized. For this reason the conditions of 
coherent neutrino production and detection are sometimes called the 
localization conditions.

Coherence of neutrino production and detection in the configuration-space 
formulation was also considered in Refs.~\cite{fzsm,paradoxes,nonrel}. 
Here I discuss it following Ref.~\cite{nonrel}. 

The 4-coordinate of the neutrino production point has an intrinsic uncertainty 
$(\delta t,\,\delta\vec{x})$ related to the finite space-time extension of the 
production process. 
This leads to the fluctuations 
$\delta\phi_{osc}\equiv \delta(\Delta\phi)$ 
of the oscillation phase (\ref{eq:phase1}): 
\be
\delta \phi_{osc}
=\Delta E \cdot \delta t - \Delta
\vec{p}\cdot
\delta\vec{x}\,. 
\label{eq:fluc1}
\ee
For neutrino oscillations to be observable, these fluctuations must be small 
-- otherwise the oscillations will be washed out upon averaging of the 
oscillation phase over the 4-coordinate of neutrino production. That is, 
a necessary condition for the observability of neutrino oscillations is  
\be
|\Delta E \cdot \delta t - \Delta\vec{p}\cdot\delta\vec{x}|\ll 1\,.
\label{eq:fluc2}
\ee
Barring accidental cancellations between the two terms in (\ref{eq:fluc2}), 
one can rewrite it as 
\be
|\Delta E\cdot\delta t| \ll 1\,,\qquad\quad |\Delta\vec{p}\cdot\delta\vec{x}|
\ll 1\,.
\label{eq:cond4}
\ee
Now, the fluctuations of the neutrino emission time and coordinate are limited 
by the temporal extension of the production process and the spatial size of 
the production region: 
\be
\delta t\lesssim \sigma_t\,,\quad\qquad |\delta\vec{x}|\lesssim \sigma_X\,.
\label{eq:cond5}
\ee
Taking into account Eq.~(\ref{eq:uncert}), from (\ref{eq:cond4}) 
we therefore obtain 
\be
|\Delta E| \ll \sigma_E\,,\quad\quad |\Delta p| \ll \sigma_p\,. 
\label{eq:cohprod}
\ee
These conditions actually have a simple meaning: 
Different neutrino mass eigenstates
can be emitted coherently and compose a flavour state only if their intrinsic 
QM energy and momentum uncertainties, $\sigma_E$ and $\sigma_p$, are
sufficiently large to accommodate their differing energies and momenta.
Similar considerations apply to neutrino detection: energy and momentum 
uncertainties related to the space-time localization of the detection 
process must be large enough to preclude determination of the neutrino 
mass, or else the oscillations will be unobservable. If by $\sigma_E$ we 
understand the smallest between of the energy uncertainties 
associated with neutrino production and detection (and similarly 
for the momentum uncertainty $\sigma_p$), conditions~(\ref{eq:cohprod}) 
will ensure coherence of both neutrino production and 
detection processes. Note that these conditions 
coincide with those in Eqs.~(\ref{eq:cond2}) and~(\ref{eq:cond3}) which 
allowed us to obtain the standard oscillation phase from the general 
expression (\ref{eq:phase1}). 

While the coherent production condition in Eq.~(\ref{eq:fluc2}) is obviously 
Lorentz invariant,\footnote{Indeed, both the oscillation phase and its
variation, being products of two 4-vectors, are Lorentz invariant.} 
the conditions in Eq.~(\ref{eq:cohprod}) are not. They follow 
from~(\ref{eq:fluc2}) only in the absence of cancellations between the 
$\Delta E \cdot \delta t$ and $\Delta\vec{p}\cdot\delta\vec{x}$ terms.  
It can be shown that even if this no-cancellation requirement is 
met in a reference frame $K$, it may be violated in 
reference frames moving with the speed $u\approx 1$ with respect to $K$ 
provided that $1-u$ is small enough~\cite{nonrel}.  
In such frames the inequalities in Eq.~(\ref{eq:cohprod}) do not play the 
role of the coherent production/detection conditions, and Eq.~(\ref{eq:fluc2}) 
should be used instead. In what follows I will be assuming that the 
no-cancellation condition is met, so that the inequalities in 
Eq.~(\ref{eq:cohprod}) do play the role of conditions for coherent neutrino 
production and detection.

\subsection{\label{sec:propdecoh}Propagation coherence and decoherence}

For neutrino oscillations to be observable it is not sufficient that the 
neutrino production and detection processes be coherent:  
In addition, coherence must not be (irreversibly) lost during neutrino 
propagation from its source to the detector. 

How could a loss of neutrino coherence in transit from the source do the 
detector occur? This question was first studied by Nussinov in what appears 
to be the first publication on the wave packet approach to neutrino 
oscillations~\cite{nussinov}.
The wave packets of the different neutrino mass eigenstates 
composing a produced flavour eigenstate propagate with different group 
velocities. This is because these velocities, $\vec{v}_{gi}\equiv
\partial E_i/\partial \vec{p}_i=\vec{p}_i/E_i$, depend on the neutrino mass 
$m_i$. Due to the difference of the group velocities $\Delta v_g$, 
over a time $t$ the centers of the wave packets of the different neutrino 
mass eigenstates separate by the distance $\Delta v_g t$. Once this distance 
exceeds the spatial length of the wave packets $\sigma_x$, 
the wave packets of different neutrino mass eigenstates cease to overlap and 
so lose their coherence. In this case neutrino oscillations cannot be 
observed. This can be seen from the fact that, once the wave packets of the 
different mass eigenstates have separated in space, one can in principle 
discriminate between them at detection, e.g., by making use of the 
time-of-flight measurement technique. 

The coherence time and coherence length can be found from the conditions 
\be
\Delta v_g \cdot t_{\rm coh}~\simeq~\sigma_x \,;\qquad\quad
l_{\rm coh}~\simeq~v_g t_{\rm coh}\,,	
\label{eq:lcoh}
\ee
where in the second equality 
$v_g$ denotes the average group velocity of the mass eigenstates 
(recall that we are assuming $\Delta m^2\ll E^2$, so that $\Delta v_g\ll v_g$). 
For ultra-relativistic neutrinos 
$\Delta v_g
\simeq \frac{\Delta m^2}{2 E^2}$, 
and Eq.~(\ref{eq:lcoh}) \vspace*{-2mm}yields~\cite{nussinov} 
\be
l_{\rm coh}\simeq \frac{v_g}{|\Delta v_g|} \sigma_x \simeq 
\frac{2 E^2}{|\Delta m^2|}\,\sigma_x\,.
\label{eq:lcoh2}
\ee
Neutrino oscillations can only be observed at the distances $L$ 
from the neutrino source satisfying $L\ll l_{\rm coh}$. 
Although the lengths of the neutrino wave packets $\sigma_x$ are usually 
microscopically small, the coherence length $l_{\rm coh}$ is 
macroscopic and very large because of the huge factor 
$2E^2/\Delta m^2$ multiplying $\sigma_x$ in the expression for $l_{\rm coh}$. 

An interesting point, first made by Kiers, Nussinov and 
Weiss~\cite{kiers1,kiers2}, is that even if the wave packets of the different 
neutrino mass eigenstates composing an emitted flavour state have separated 
on their way  between the neutrino source and detector, their coherence 
may be restored at neutrino detection. The point is that any detection process 
is not instantaneous -- it takes a finite time $\sigma_{t\,det}$ (which is 
related to the ultimate energy resolution of detection $\sigma_{E\,det}$ by 
$\sigma_{t\,det}\sim \sigma_{E\,det}^{-1}$). Assume that the elementary 
detection process lasts long enough, so that the wave packets of the different 
neutrino mass eigenstates, which have separated during the neutrino 
propagation, arrive at the detector before the detection is over. Then their 
detection amplitudes may add coherently and interfere, leading to observable 
flavour oscillations. Possible restoration of propagation coherence at 
detection can be automatically taken into account if in the expression for the 
coherence length in Eq.~(\ref{eq:lcoh2}) by $\sigma_x$ we understand an 
``effective'' length of the neutrino wave packet, defined as 
$\sigma_x=v_g/\sigma_E$ with $\sigma_E$ being the smaller between 
the QM energy uncertainties associated with neutrino production and detection. 
In particular, in the limit $\sigma_{E\,det}\to 0$ the coherence length 
formally becomes infinite, i.e.\ decoherence by wave packet separation never 
occurs. Note, however, that for too small $\sigma_{E\,det}$ the condition of 
coherent detection of different neutrino mass eigenstates (\ref{eq:cohprod}) 
may be violated. So, the issue of compatibility of the different coherence 
conditions should be considered.

\subsection{\label{sec:compat}Are different coherence constraints compatible?}

We have found that there are two types of coherence conditions that have to be 
satisfied in order for neutrino oscillations to be observable: (i) coherence 
of neutrino production and detection and (ii) coherence of neutrino 
propagation. Before proceeding to discuss their compatibility, 
let me make the following point. 
One can show that under very general assumptions the second condition in 
(\ref{eq:cohprod}) follows from the first one and so is actually 
redundant~\cite{nonrel}.  That is, the first condition in (\ref{eq:cohprod}) 
by itself ensures coherence of neutrino production and detection. 

The production/detection and the 
propagation coherence conditions, $\Delta E\ll \sigma_E$ and $L\ll 
l_{\rm coh}$, both put upper limits on the neutrino mass squared 
difference $\Delta m^2$:  
\be
\Delta E\sim \frac{\Delta m^2}{2E}\ll \sigma_E\,,\qquad 
\frac{\Delta m^2}{2E^2}L \ll \sigma_x\simeq v_g/\sigma_E\,.
\vspace*{-1mm}
\label{eq:compat1}
\ee
However, when expressed as constraints on $\sigma_E$, they read 
(taking into account that $v_g\approx 1$)  
\be
\frac{\Delta m^2}{2E}\ll \sigma_E \ll \frac{2E^2}{\Delta m^2}\frac{1}{L}\,, 
\label{eq:compat2}
\ee
that is, they constrain $\sigma_E$ both from above and from below. A natural 
question then is: Are these constraints compatible with each other? With 
decreasing $\Delta m^2$ the left hand side (l.h.s.) of Eq.~(\ref{eq:compat2}) 
decreases, while its r.h.s.\ increases; so it is clear that the smaller 
$\Delta m^2$, the easier it is to satisfy the conditions in 
Eq.~(\ref{eq:compat2}).\,\footnote{
By the way, this may already give us a hint on what the answer 
to the question ``Do charged leptons oscillate?'' should be.} 
For the two conditions in Eq.~(\ref{eq:compat2}) to be compatible, its 
l.h.s.\ must be small compared to its r.h.s., which gives 
\be
2\pi \frac{L}{l_{\rm osc}} ~\ll~
\frac{2E^2}{\Delta m^2}\;\;(\gg 1)\,. 
\label{eq:compat3}
\ee
It should be stressed that this condition is necessary but in general not 
sufficient for a mixed neutrino state to be coherently produced, maintain 
its coherence over the distance $L$ and then be coherently detected: it only 
ensures the consistency of the two conditions in  
Eq.~(\ref{eq:compat2}) but not their separate fulfilment.
Note that from the rightmost inequality in Eq.~(\ref{eq:compat2}) it follows 
that maximum number of observable oscillations $(L/l_{\rm osc})_{\rm max}$  
cannot exceed
\be
\frac{l_{\rm coh}}{l_{\rm osc}}\sim 
\frac{1}{2\pi}\frac{E}{\sigma_E}\,.
\label{eq:compat4}
\ee
In reality, the observable number of oscillations 
is much smaller: it is obtained from the r.h.s.\ of  Eq.~(\ref{eq:compat4}) 
by replacing $\sigma_E\to \delta E$, where $\delta E$ is the energy 
resolution of the detector which is usually much larger than the ultimate 
QM energy uncertainty $\sigma_E$. This puts an upper limit on the baselines at 
which the oscillations can be observed. For longer baselines, the flavor 
transition (and survival) probabilities 
will correspond to Eq.~(\ref{eq:master1}) with all the oscillatory terms 
replaced by their averaged values. 

\subsection{\label{sec:really}Are they actually satisfied?}

Are the coherence conditions normally satisfied in neutrino oscillation 
experiments? For oscillations between the usual flavour-eigenstate neutrinos 
$\nu_e$, $\nu_\mu$ and $\nu_\tau$ in the 3-flavour scheme, the coherent 
production and detection conditions are satisfied with a large margin  
in all cases of practical interest. This is a consequence of the tininess of 
the masses of their mass-eigenstate counterparts $\nu_1$, $\nu_2$ and $\nu_3$ 
and is now firmly established by the positive results of the experiments on 
atmospheric, reactor and accelerator neutrino oscillations. 

A rather obvious though not widely recognized is the fact that even 
non-observation of neutrino flavour transitions in experiments with the 
detector situated too close to the neutrino source (as it was the case 
e.g.\ in `prehistoric' reactor and accelerator neutrino experiments) is a 
direct consequence of and a firm evidence for coherence of neutrino production 
and detection. Indeed, if coherence was violated, 
i.e., if different neutrino mass eigenstates were emitted or absorbed 
incoherently, one would have observed a suppression of the original neutrino 
flux (in the disappearance experiments) or an emergence of ``wrong-flavour'' 
neutrinos (in the appearance experiments) corresponding to the averaged 
oscillation probabilities.\,\footnote{For example, in the 2-flavour case the 
survival probability at short baselines $L\ll l_{\rm osc}$ is 
$P_{\alpha\alpha}=(\sin^2\theta+\cos^2\theta)^2=1$ in the coherent case 
and $P_{\alpha\alpha}=\sin^4\theta+\cos^4\theta<1$ if coherence is strongly 
violated at neutrino production or detection.}  
As the leptonic mixing angles are relatively large (especially $\theta_{23}$ 
and $\theta_{12}$), effects of such flavour transitions would have been quite 
sizeable. 

How about the propagation coherence condition? For 3-flavour oscillations 
between $\nu_e$, $\nu_\mu$ and $\nu_\tau$ it is satisfied with a large margin 
for all terrestrial experiments. In particular, from the atmospheric neutrino 
experiments we know that the coherence holds over macroscopic distances as 
large as about 10,000 km. At the same time, for astrophysical and cosmological 
neutrinos which propagate enormous distances before reaching the earth, 
coherence is lost. Thus, solar and supernova neutrinos arrive at terrestrial 
detectors as incoherent mixtures of the mass eigenstates rather than as 
flavour eigenstate neutrinos. Cosmic background neutrinos should also at 
present be in mass rather than in flavour states. 

The situation with coherence of neutrino flavour transitions may, however, be 
different if relatively heavy predominantly sterile neutrinos exist. Note 
that heavy neutrinos with masses in the eV -- keV -- MeV (and even GeV) 
ranges are now being actively discussed in connection with some hints from 
short-baseline accelerator experiments (LSND, MiniBooNE), reactor neutrino 
anomaly, anomaly in gallium radioactive source experiments, keV sterile 
neutrinos as dark matter, pulsar kicks, leptogenesis via neutrino oscillations, 
supernova $r$-process nucleosynthesis, unconventional contributions to 
$2\beta 0\nu$ decay, etc.~\cite{ster}. For the corresponding large mass 
squared differences the production/detection and propagation coherence 
conditions may be violated, leading to important implications. Therefore in 
situations when large $\Delta m^2$ may play a role the fulfilment of the 
coherence  conditions should be carefully examined in each particular case. 

\section{\label{sec:WP2}Wave packet approach: Gaussian wave packets}

So far I have been discussing neutrino production/detection and propagation 
coherence as necessary conditions for observability of neutrino oscillations 
in a rather qualitative way, basing on some very general arguments. Can we see 
how this works in an explicit calculation? As I discussed earlier, for 
this one needs to resort to wave packets. 

To the best of my knowledge, the first complete calculation of the oscillation 
probability in the wave packet framework was performed by Giunti, Kim and 
Lee~\cite{gkl1,gkl2}, though simplified studies can also be found in some 
earlier papers (see Section \ref{sec:hist} below). Describing neutrinos by 
Gaussian wave packets, Giunti et al.\ found the following expression for the 
probability of $\nu_\alpha\to\nu_\beta$ oscillations: 
\be
P_{\alpha\beta}(L,E)=\sum\limits_{i,k}U_{\alpha i}U^*_{\beta i}U^*_{\alpha k}
U_{\beta k} e^{-i (\Delta m_{ik}^2/2E)L}
\,e^{-[\Delta E_{ik}^2/8\sigma_E^2]-[L/(l_{\rm coh})_{ik}]^2}\,.
\vspace*{-1.5mm}
\label{eq:gauss}
\ee
Here the indices $i,k$ correspond to neutrino mass eigenstates and 
\be
(l_{\rm coh})_{ik}
=2\sqrt{2}\frac{2E^2}{|\Delta m_{ik}^2|}\sigma_x\,
\label{eq:lcoh3}
\ee
is the coherence length (which is to be compared with Eq.~(\ref{eq:lcoh2}) 
found from qualitative arguments). The quantity $\sigma_E$ is the effective 
neutrino energy uncertainty which is expressed through the energy 
uncertainties related to neutrino production and detection as 
\be
\frac{1}{\sigma_E^2}=\frac{1}{\sigma_{E\,prod}^2}+
\frac{1}{\sigma_{E\,det}^2}\,,
\label{eq:eff}
\ee
i.e.\ it is dominated by the smaller of the two. The last exponential factor 
on the r.h.s.\ of Eq.~(\ref{eq:gauss}), 
\be
e^{-[\Delta E_{ik}^2/8\sigma_E^2]-[L/(l_{\rm coh})_{ik}]^2}\,,
\label{eq:factor}
\ee
takes into account possible decoherence effects. It strongly suppresses the 
off-diagonal terms in the summand of Eq.~(\ref{eq:gauss}) when either 
$|\Delta E_{ik}|\gg \sigma_E$, which means violation of production or 
detection coherence, or $L\gg (l_{\rm coh})_{ik}$, which implies 
(irreversible) decoherence by wave packet separation. The suppression of the 
off-diagonal terms in (\ref{eq:gauss}) would mean that all the oscillatory 
terms in the oscillation probability are effectively replaced by their 
averages, giving 
\be
P_{\alpha\beta}(L,E)\to \bar{P}_{\alpha\beta}\equiv  
\sum\limits_{i}|U_{\alpha i}|^2 |U_{\beta i}|^2\,,
\vspace*{-2mm}
\label{eq:average}
\ee
which is $L$- and $E$-independent. 
Note that the same result is obtained from the standard expression for 
the oscillation probability in Eq.~(\ref{eq:master1}) upon averaging out  
all its oscillatory terms.

On the other hand, if the production/detection and propagation coherence 
conditions are satisfied for all $(i,k)$, one can replace the last 
exponential factor in Eq.~(\ref{eq:gauss}) by unity; the resulting 
probability of flavour transitions then coincides with the standard 
probability of neutrino oscillations in vacuum 
(\ref{eq:master1}).\,\footnote{Here and in most of the following text  
I neglect the difference between $p$ and $E$ which is justified for 
ultra-relativistic neutrinos. The exception is Section~\ref{sec:lorentz} 
where the Lorentz invariance issues are discussed.}

\section{\label{sec:norm}Wave packet approach and the normalization problem} 

In deriving the oscillation probabilities in the wave packet approach, 
one encounters the following problem: by using the standard normalization of 
the wave packets in the coordinate space $\int d^3x |\Psi_i(t,\,\vec{x})|^2=1$ 
(or equivalently, in momentum space, 
$\int[d^3p/(2\pi)^3] |f_{\vec{p}_0}(\vec{p})|^2=1$) one does not get the 
correct normalization for the oscillation probability. Instead, the result 
differs from the standard oscillation probability by a constant factor. 
The usual way to circumvent this difficulty is to introduce arbitrary 
normalization factors for the wave functions of the produced and detected 
neutrino states and then fix them at the end of the calculation by imposing 
on the oscillation probabilities the unitarity condition 
\be
\sum_\beta P_{\alpha\beta}(L,E)=1\,.
\vspace*{-2mm}
\label{eq:unitarity}
\ee 
Although it works, this is an {\em ad hoc} procedure which looks rather 
unsatisfactory. Can one avoid it by using a different normalization of the 
neutrino wave functions? 
 
It turns out that the answer to this question is negative: no separate 
(i.e.\ independent) normalization of 
the wave functions of the produced and detected neutrino states would 
result in the correctly normalized $P_{\alpha\beta}(L,E)$. 
For example, if $f^P_{\vec{p}_0}(\vec{p})$ and $f^D_{\vec{p}_0}(\vec{p})$ are 
the momentum-space wave functions of the produced and detected neutrinos 
(with the indices $P$ and $D$ standing for production and detection), 
the correct normalization of the oscillation probability is only obtained 
provided the condition 
\be
\int\frac{d^3p}{(2\pi)^3}|f^P_{\vec{p}_0}(\vec{p})|^2
|f^D_{\vec{p}_0}(\vec{p})|^2=1\,
\label{eq:overlap}
\ee
is satisfied. That is, in order to obtain the correctly normalized oscillation 
probability, one needs to impose a correlated normalization condition for 
the wave functions of the produced and detected neutrino states; no separate 
normalization of them would do the job.   

It is, actually, not difficult to understand why this happens. The problem is 
related to the wave packet description of the process itself and not to 
neutrino oscillations: we would have encountered a similar problem when 
considering neutrino production and subsequent detection in the wave 
packet picture even if there existed just one neutrino species in Nature, 
i.e.\ if the oscillations were absent. The point is that in the wave packet 
approach each individual particle (and not just an ensemble of them) is 
characterized by a spectrum of momenta (and energies). For the produced and 
detected neutrinos, the corresponding spectra are dictated by the character and 
the properties of the production and detection processes, such as their 
localization. The production and detection processes are different, and so are 
the corresponding momentum spectra. The detection efficiency therefore depends 
on the degree of overlap of these spectra, which is given by the integral on 
the l.h.s.\ of Eq.~(\ref{eq:overlap}). This just reflects energy and momentum 
conservation. In particular, in the special case when the overlap is absent, 
i.e.\ when the momentum modes corresponding to detection are absent from the 
spectrum of the produced wave packets (which happens when the energy threshold 
of the detection process is above the maximum energy of the produced 
particles), there will be no detection at all. 

Coming back to neutrino oscillations, 
kinematic prohibition of neutrino detection would not, of course, mean that 
neutrinos do not oscillate. On the other hand, even if the overlap of the 
spectra of the produced and detected neutrinos is perfect, i.e.\ the integral 
on the l.h.s.\ of Eq.~(\ref{eq:overlap}) takes its maximum possible value,  
this does not solve the normalization problem~\cite{AkhKopp}. The point is that 
the overlap integral is simply not a part of the oscillation probability. 
Obviously, the oscillation probability itself should be independent of the 
efficiency of neutrino detection. In calculating $P_{\alpha\beta}(L,E)$, 
all the details of the neutrino production and detection processes should be 
factored out and removed. Normalization of $P_{\alpha\beta}(L,E)$ by imposing 
the unitarity constraint does just this. I will discuss this point in more 
detail in the next Section, where the related issue of universality of the 
oscillation probabilities is considered.

\section{\label{sec:universal}Universal oscillation probabilities?} 

The standard oscillation probability (\ref{eq:master1}) depends only on 
neutrino energy $E$ and the distance from the source $L$, but not on the 
processes in which the neutrinos were produced and detected. 
Under what conditions is this justified? That is, when are neutrino 
oscillations actually described by a universal (i.e.\ production-- and 
detection--independent) probability? 

Strictly speaking, in general one should consider neutrino production, 
propagation and detection as a single process, as only the probability of the 
complete process, $\Gamma_{\alpha\beta}(L, E)$, is directly related to 
measurable quantities in oscillation experiments. Under certain conditions 
$\Gamma_{\alpha\beta}(L, E)$ can be factorized as 
\be
\Gamma_{\alpha\beta}(L, E)=j_\alpha(E)\,
P^{\rm prop}_{\alpha\beta}(L,E)\sigma_\beta(E)\,,
\label{eq:factoriz1}
\ee
where $j_\alpha(E)$ is the flux of the produced $\nu_\alpha$, 
$P^{\rm prop}_{\alpha\beta}(L,E)$ is the probability of neutrino propagation 
between its source and detector, which takes into account possible 
$\nu_\alpha\to\nu_\beta$ transitions and also includes a  
geometrical factor describing neutrino flux suppression with increasing 
distance $L$ from the source, and $\sigma_\beta(E)$ is the detection cross 
section for $\nu_\beta$. If such a factorization is possible, one can find 
the oscillation probability $P_{\alpha\beta}(L,E)$ by dividing 
$\Gamma_{\alpha\beta}(L, E)$ by $j_\alpha(E)$, $\sigma_\beta(E)$ and by the 
trivial geometric factor of neutrino flux suppression with distance.\footnote{
Note that this procedure will automatically lead to the correctly normalized 
oscillation probability 
because it rids the transition probability of the details of the neutrino 
production and detection processes.} If, however, the factorization 
(\ref{eq:factoriz1}) turns out to be impossible, the production-- and 
detection--independent oscillation probability cannot even be defined. In this 
case one should instead consider the probability $\Gamma_{\alpha\beta}(L,E)$ 
of the overall neutrino production--propagation--detection process. 

So, when is the factorization (\ref{eq:factoriz1}) actually possible? 
It can be shown~\cite{paradoxes,AkhKopp} that for such a factorization 
to take place the following conditions must be satisfied:

\begin{itemize} 

\item[(a)]
Neutrinos are ultra-relativistic or quasi-degenerate in mass;  

\item[(b)]
Neutrino production, propagation and detection processes are coherent, i.e.\ 
they do not allow one to discriminate 
between different neutrino mass eigenstates. 

\end{itemize} 
These conditions can be easily understood. The factorization 
(\ref{eq:factoriz1}) only takes place when all the three processes -- neutrino 
production, propagation and detection -- are independent of each other. 
If condition (a) is violated, the composition of the produced neutrino state 
$\nu_\alpha$ in terms of the mass eigenstates $\nu_i$ will not be given by 
Eq.~(\ref{eq:mix}), but will rather depend sensitively on the neutrino masses 
$m_i$ in a way that depends on the kinematics of the production process.
As the flavour transition probabilities depend on the composition of 
the produced neutrino state, neutrino production and propagation processes 
will not be independent in this case. If condition (b) is not met, the 
probabilities of flavour transformations will depend on the degree of 
coherence violation at neutrino production and/or detection, so that there will 
again be no independence of neutrino production, propagation and detection. 

If we compare conditions (a) and (b) for the universal oscillation probability 
to exist with the discussed earlier conditions for neutrino oscillations in 
vacuum to be described by the standard oscillation probability, we will find 
that they exactly coincide. This is an important point: {\sl whenever the 
universal (production-- and detection--independent) oscillation probability 
makes sense at all, it is given by the standard oscillation 
formula~(\ref{eq:master1}).}
  
What happens if the coherence conditions are partially or completely violated? 
In this case one can in principle introduce an effective oscillation 
probability by {\em defining} it as the probability of the overall process 
$\Gamma_{\alpha\beta}(L, E)$ divided by the flux of the produced neutrinos 
$j_\alpha(E)$, detection cross section $\sigma_\beta(E)$ and the geometric 
factor of neutrino flux suppression with distance. 
Such an effective oscillation probability will obviously be non-universal. 
It can be shown that, if condition (a) above is satisfied and in addition 
either neutrino production or detection is coherent, the so defined effective 
oscillation probability will be correctly normalized. An example of such 
a production-- and detection--dependent effective oscillation probability 
is given by Eq.~(\ref{eq:gauss}).    

More detailed discussion of the conditions for the existence of the universal 
and/or correctly normalized oscillation probability can be found in 
Ref.~\cite{AkhKopp}. 

\section{\label{sec:corr}Small corrections?}
We know that the standard formula for the oscillation probability 
(\ref{eq:master1}) 
is correct and works well in most cases of practical interest, provided that 
matter effects on neutrino oscillations are absent or can be neglected. 
However, it is obviously not exact, and so one may be tempted to look for 
(presumably) small corrections to it. For example, one might look for the 
corrections to the standard oscillation phase due to the next terms in the 
relativistic expansion of neutrino energy $(p_i^2+m_i^2)^{1/2}\approx 
p_i +m_i^2/(2p_i)+...$\,. The next-to-leading order contribution to the 
oscillation phase $\phi_{\rm osc}$ will then be proportional to $\Delta m^4$. 
At first sight, this may make sense. However, the correction will only become 
noticeable at baselines $L$ at which its contribution to the oscillation phase 
becomes comparable to 1. It is easy to see that for such distances the leading 
order term in the phase is of order $E^2/m_\nu^2\ggg 1$; this means that 
the oscillations are already in the complete averaging regime, and any  
correction of order one to the oscillation phase are irrelevant.

\section{\label{sec:conserv}Are neutrino oscillations compatible with 
energy-momentum conservation?} 

In quantum theory, the rates of processes are calculated by making use of the 
generalized Fermi's golden rule 
\be
\Gamma~=~\prod_i\frac{1}{(2E_i)}\int \prod_f \frac{d^3 p_f}{(2\pi)^3\, 2E_f}
\big|M_{fi}\big|^2 (2\pi)^4\delta^4\Big(\sum_f p_f - \sum_i p_i\Big)\,,
\label{eq:fermiGL}
\ee
where the factor $\delta^4(\sum_f p_f - \sum_i p_i)$ ensures energy-momentum 
conservation. In particular, it is used to calculate neutrino production rates 
and detection cross sections. However, if one applies it e.g.\ to neutrino 
production, one might conclude that the neutrino 4-momentum $p=(E,\,\vec{p}\,)$ 
can be determined from the 4-momenta of all other particles participating in 
the production process. But then from the on-shell relation $E^2=\vec{p}\,^2+
m^2$ one would be able to find the neutrino mass, which would mean that the 
produced neutrino is a mass eigenstate and not a flavour one. This would 
imply that neutrinos cannot oscillate. Similar arguments could be made for 
neutrino detection. 

Thus, a dichotomy arises: 
On the one hand, energy-momentum conservation is, to the best of our knowledge, 
an exact law of nature. On the other hand, applying energy and momentum 
conservation to neutrino production or detection would apparently make 
neutrino oscillations impossible, in contradiction with experiment. 

This caused a significant confusion in the literature. As I discussed above, 
the resolution of the paradox comes from the observation that particles 
participating in neutrino production and detection processes are localized in 
space and time, and therefore their energies and momenta have intrinsic QM 
uncertainties. Although these uncertainties are usually very small, they 
cannot be ignored when neutrino oscillation are considered. In other words, 
one has to take into account that the states of these particles are not exact 
eigenstates of energy and momentum. This does not mean, of course, that 
energy and momentum conservation laws are violated. 

\section{\label{sec:lorentz}\,Lorentz invariance of the oscillation 
probabilities} 

The probabilities of neutrino flavour transitions must not depend on the 
Lorentz frame in which the oscillations are considered, i.e.\ must be Lorentz 
invariant. Can we see that this is indeed the case, in particular, is the 
standard oscillation probability (\ref{eq:master1}) invariant with respect to 
Lorentz transformations?

In addition to its dependence on the neutrino mass squared differences and the 
parameters of the leptonic mixing matrix $U$ which are universal constants, 
the expression in eq. (\ref{eq:master1}) depends only on the distance $L$ 
from the neutrino source and the mean momentum of the neutrino state $p$ 
through their ratio $L/p$. Is this quantity Lorentz invariant? It is not 
difficult to show that it is, provided that the condition $L=v_gt$ is 
satisfied~\cite{levy}. Note that the relation $L=v_gt$ is itself Lorentz 
covariant \cite{paradoxes}; it essentially means that neutrinos are considered 
as point-like particles. I criticized the use of this approximation within the 
plane-wave approach, but it is perfectly legitimate to employ it in the wave 
packet framework provided that the length $\sigma_x$ of the wave packet is 
small compared to the other characteristic lengths inherent to the problem. 
In the case of neutrino oscillations it is justified when $\sigma_x$ is 
vanishingly small compared to the neutrino oscillation length. It can be shown 
that this requirement is indeed met when the coherent neutrino production and 
detection conditions are fulfilled~\cite{nonrel}. 
As $L/p$ is Lorentz invariant in this limit, so 
is the standard oscillation probability~(\ref{eq:master1}).\,\footnote{In the 
literature it is often stated that the neutrino oscillation probabilities 
in vacuum depend on $L/E$. However, careful derivation in the wave packet 
framework actually yields the dependence on $L/p$. While for relativistic 
neutrinos $L/E$ is essentially the same as $L/p$, when considering the Lorentz 
invariance issues it is important to remember that $L/p$ is Lorentz invariant 
(provided that $L=v_g t$ holds), whereas $L/E$ is not.} 
A more detailed discussion of the Lorentz invariance of the neutrino 
oscillation probability in the wave packet approach, including the case when 
the coherence conditions are not satisfied, can be found in~\cite{paradoxes}. 

\section{\label{sec:charged}\,Do charged leptons oscillate?} 

The Lagrangian of charged-current leptonic weak interactions is completely 
symmetric with respect to charged leptons and neutrinos, so why do we say 
that the charged leptons are produced and absorbed in these interactions as 
mass eigenstates ($e$, $\mu$, $\tau$), whereas neutrinos as flavour eigenstates 
(\ref{eq:mix}) which are linear superpositions of the mass eigenstates 
$\nu_1$, $\nu_2$ and $\nu_3$? Why not the other way around? Or why aren't 
both charged leptons and neutrinos produced and detected as linear 
superpositions of their respective mass eigenstates?  
After all, the mixing matrix $U$ comes from the diagonalization of both 
neutrino and charged lepton mass matrices, so it is the {\em leptonic} mixing 
matrix (and not ``the neutrino mixing matrix'', as it is sometimes incorrectly 
called). A related question is: do charged leptons oscillate? 

We know that neutrinos are emitted and detected as coherent 
linear superpositions of different neutrino mass eigenstates only when 
the coherence conditions for their production, propagation and detection 
are satisfied, and that all these conditions put upper limits on the 
neutrino mass squared differences $\Delta m^2$. Similar considerations 
apply also to charged leptons; the difference is that their mass 
squared differences are many orders of magnitude larger than those of 
neutrinos. As a result, for the charged leptons the coherence conditions 
are not satisfied -- these particles are always produced and detected as mass 
eigenstates and not as coherent superpositions of the mass eigenstates. This, 
in particular, means that they do not oscillate~\cite{charged}. 

This also tells us that neutrinos produced, e.g., in $\pi\to\mu\nu$ and 
$\pi\to e\nu$ decays oscillate even when the corresponding charged leptons 
are not detected. For neutrino oscillations to take place, the initially 
produced neutrino state has to be a flavour eigenstate -- a well-defined 
coherent superposition of the neutrino mass eigenstates. The absence of 
coherence of different charged leptons ensures that in each charged pion 
decay event either $\mu$ or $e$ is produced but not their linear 
superposition. This provides a measurement of the flavour of the associated 
neutrino, which is necessary for neutrino oscillations to occur~\cite{charged}. 

\section{\label{sec:hist}\,A bit more history}
The literature on QM aspects of neutrino oscillations and, in particular, on 
their wave-packet description, is vast. I have already mentioned several 
papers before; I will now discuss them in a bit more detail and will also give 
a very brief overview of a few more. More detailed discussions and further 
references can be found in~\cite{zralek,beuthe,naumov}.  

As mentioned above, neutrino oscillations were first considered in the wave 
packet approach by Nussinov~\cite{nussinov}. He discussed the effects of 
decoherence by wave packet separation and pointed out the existence of the 
coherence length $l_{\rm coh}\approx \sigma_x(v_g/|\Delta v_g|)$. He also 
estimated the lengths of the neutrino wave packets $\sigma_x$ and the coherence 
lengths $l_{\rm coh}$ for neutrinos produced in decays of isolated 
nuclei ($\sigma_x\approx c/\Gamma$, where $c$ is the speed of light 
and $\Gamma$ is the decay width of the parent particle), as well as for 
neutrinos from nuclear beta decay in the interior of the sun. For the 
lengths of the wave packets of solar neutrinos he found $\sigma_x\approx 
c\tau$, where $\tau$ is the time of un-interrupted neutrino emission by 
the nucleus, which essentially coincides with the time between collisions 
significantly changing the phase of the emitter. Assuming $\rho\sim 
100~{\rm g/cm^3}$ and $T\sim 1~{\rm keV}$ as the typical density and 
temperature in the solar core, for solar $^7$Be neutrinos he found 
$\tau\sim 3\times 10^{-17}\,{\rm s}$ and $l_{\rm coh}\sim 20~{\rm km}$. In his 
estimate of the coherence length he used $\Delta m^2\sim 1$~eV$^2$; with the 
currently known value of this quantity he would have obtained 
$l_{\rm coh}\sim 3\times 10^5~{\rm km}$.  

As discussed in Section \ref{sec:cohCond}, Kayser~\cite{kayser} considered 
the coherence conditions for neutrino production and detection and related 
them to the space-time localization of these processes. He also presented a 
simplified analytic description of neutrino oscillations in the wave packet 
picture.

Kobzarev, Martemyanov, Okun and Shchepkin~\cite{kmos} were the first to  
include neutrino production and detection processes in the analysis of 
neutrino oscillations. They used a simplified model in which the neutrino 
source and detected were assumed to be infinitely heavy. 
 
The first complete derivation of the neutrino oscillation probability within 
the wave packet approach was given by Giunti, Kim and Lee~\cite{gkl1,gkl2}. 
They described neutrinos by Gaussian wave packets and explicitly demonstrated 
how the oscillations get suppressed when coherence conditions are violated. 

In Ref.~\cite{gkl3} Giunti, Kim, Lee and Lee included neutrino production and 
detection processes with the source and target particles localized and 
described by Gaussian wave packets. A similar approach was independently 
developed by Rich~\cite{rich}.  

Kiers, Nussinov and Weiss~\cite{kiers1,kiers2} pointed out that neutrino 
coherence lost on the way between the source and detector due to wave packet 
separation can in principle be recovered at detection. In Ref.~\cite{kiers2} 
they also considered neutrino production, propagation and detection as a 
single process in a simple model with localized neutrino source and detector. 

Farzan and Smirnov~\cite{fzsm} considered neutrino propagation decoherence in 
momentum space as the effect of accumulation with distance of fluctuations of 
the oscillation phase due to the momentum spread within the wave packet. They 
also demonstrated Lorentz invariance of the product $\sigma_x E$ and, on a 
different note, pointed out that the spatial spreading of the neutrino wave 
packets does not affect neutrino oscillations. 

In Ref.~\cite{paradoxes} a shape-independent wave packet approach to neutrino 
oscillation was developed, the same energy/same momentum confusion was cleared 
up (see also Dolgov~\cite{dolgov} for an earlier discussion), and Lorentz 
invariance of the oscillation probability in both coherent and decoherent 
cases was demonstrated. 

In Ref.~\cite{AkhKopp} the QM wave packet approach to neutrino oscillations 
was compared with the one based on quantum field theoretic techniques, 
the issue of the normalization of the oscillation probability was clarified
and the conditions for the existence of a universal (production-- and 
detection--independent) oscillation probability were found. 

In Ref.~\cite{nonrel} the question of whether non-relativistic neutrinos can 
oscillate and the related Lorentz invariance issues were addressed. 

\section{\label{sec:sum}\,Summary and discussion}

Being a quantum-mechanical interference phenomenon, neutrino oscillations owe 
their very existence to the QM uncertainty relations. It is the energy and 
momentum uncertainties of neutrinos related to the space-time localization of 
their production and detection processes that allow the neutrinos 
to be emitted and absorbed as coherent superpositions of the states of well 
defined and different mass. Energy and momentum uncertainties also determine 
the lengths of the neutrino wave packets and are therefore crucial to the 
issue of the loss of coherence due to the wave packet separation.  Since 
coherence is essential for neutrino oscillations, and particles states with 
intrinsic energy and momentum uncertainties are described by wave packets, 
a consistent derivation of the oscillations probability requires using the 
wave packet formalism. 

That being said, it does not mean that each time we want to compute the 
oscillation probability 
for a neutrino experiment 
we have to resort to a full-scale wave packet 
calculation. The tininess of the neutrino mass means that we normally deal 
only with ultra-relativistic neutrinos and that in most situations the 
coherence conditions are satisfied with a large margin. Under these conditions 
the probability of flavour transitions in vacuum reduces to the well known 
standard oscillation probability (\ref{eq:master1}), which can be safely used 
most of the time provided that matter effects on neutrino oscillations are 
either absent or can be neglected. Coherence conditions, however, may be 
violated if relatively heavy predominantly sterile neutrinos exist -- in that 
case their fulfilment has to be checked on the case-by-case basis.

The standard formula for the probability of neutrino oscillations 
in vacuum (\ref{eq:master1}) is stubbornly robust -- it is not easy to find 
a situation in which it does not work or needs significant corrections. 
In addition to being perfectly accurate for ultra-relativistic neutrinos in 
the cases when the production/detection and propagation coherence conditions 
are satisfied, it can also be utilized when coherence is strongly violated 
-- one simply has to replace all the oscillatory terms in 
Eq.~(\ref{eq:master1}) by their averages. Significant deviations from 
Eq.~(\ref{eq:master1}) can only be expected when violation of coherence is 
moderate; if such situations exist at all, they should be quite rare.  

The idea of neutrino oscillations was put forward by Pontecorvo over 60 
years ago, and more than 20 years have already passed since their 
discovery. The theory of neutrino oscillations has been actively 
advanced since the 1960s and is quite mature and developed now. 
Consistent application of quantum theory allowed us to resolve numerous 
subtle issues and apparent paradoxes of the oscillation theory. In spite 
of this, attempts at revising some of its basic ingredients do not cease 
even now, usually for no good reason. Also, oversimplified approaches to 
the derivation of the oscillation probability can still be often found 
in modern reviews, lecture notes and textbooks. While the use of such 
simplifications in the pioneering papers is quite 
understandable\,\footnote{Though these simplified treatments lacked 
justification, they actually led to the correct results, which 
demonstrates a good physical intuition on the part of their authors.}, 
using them in contemporary literature can hardly be justified. 

Though quite mature, the theory of neutrino oscillations is in my opinion far 
from being closed. Over the years, many times it had appeared to us 
to be complete and finished, but each time this turned out to be wrong. 
I believe that we are still in the same situation now.

\end{document}